\documentclass[aps,pra,showpacs,amsmath,amssymb,reprint,10pt,longbibliography,noeprint,superscriptaddress]{revtex4-2}
\pdfoutput=1

\usepackage{amsmath,amsfonts,amssymb,amsthm,bbm,graphicx,enumerate,times}
\usepackage{mathtools}
\usepackage[usenames,dvipsnames]{color}
\usepackage{todonotes} %[disable]
\usepackage{comment}
\usepackage{float}
\usepackage{hyperref}
\usepackage{tikz}
\usepackage{graphicx}
\usepackage{bm}
\usetikzlibrary{shapes,positioning}
\usepackage{etoolbox}

 \newtheorem{theorem}{Theorem}

 \newtheorem{lemma}[theorem]{Lemma}

\definecolor{ingo}{rgb}{.1,.8,.2}

\definecolor{henrik}{rgb}{.8,.3,0}

\newcommand{\mc}[1]{\mathcal{#1}}

\newcommand{\mb}[1]{\mathbbm{#1}}

\newcommand{\e}{\mathrm{e}}

\newcommand{\rmd}{\mathrm{d}}
\newcommand{\Tr}{\mathrm{Tr}} %new

\newcommand{\id}{\mb{1}}

\newcommand{\vertiii}[1]{{\left\vert\kern-0.25ex\left\vert\kern-0.25ex\left\vert #1 
    \right\vert\kern-0.25ex\right\vert\kern-0.25ex\right\vert}}

\newcommand{\CC}{\mb{C}}

\newcommand{\norm}[1]{\left\Vert #1 \right\Vert}

\newcommand{\ket}[1]{|#1\rangle}

\newcommand{\bra}[1]{\langle #1|}

\newcommand{\ketbra}[2]{\ket{#1} \! \bra{#2}}

\newcommand{\sandwich}[3]
  {\left\langle  #1 \right| #2 \left| #3 \right\rangle}
  \newcommand{\proj}[1]{\ketbra{#1}{#1}}
\renewcommand{\vec}[1]{\pmb{#1}}

\newcommand*{\bE}{\mb E}

\newcommand{\fu}{Dahlem Center for Complex Quantum Systems, Freie Universit{\"a}t Berlin, 14195 Berlin, Germany}

\newcommand{\tii}{Quantum Research Centre, Technology Innovation Institute (TII), Abu Dhabi}

\begin{document}

\title{High-temperature thermalization implies the emergence of quantum state designs}
\begin{abstract}
	It was recently observed that in certain thermalizing many-body systems, measuring the complement of a subsystem that thermalized to infinite temperature in a suitable orthonormal basis gives
	rise to approximate quantum state $k$-designs as post-measurement states on the subsystem. 
	We prove that this emergence of approximate $k$-designs holds true in \emph{every} large system where some small subsystem is close to being maximally mixed.
On a technical level we show that any high-dimensional purification of an approximately maximally mixed state induces an approximate quantum state $k$-design when measured in a suitable orthonormal basis. 
	Moreover, we show that this is true with overwhelming probability for measurement bases chosen uniformly at random from the Haar measure.  
\end{abstract}
\author{Henrik Wilming}
\affiliation{Leibniz Universit\"at Hannover, Appelstra\ss e 2, 30167 Hannover, Germany}
\author{Ingo Roth}
\affiliation{\tii}
\affiliation{\fu}

\maketitle

When a complex quantum many-body system is initialized in a low-entangled state $\ket{\Psi(0)}$ and let to evolve unitarily under its Hamiltonian, we typically find that after some initial equilibration time, the quantum state $\rho_A(t) = \Tr_A[\proj{\Psi(t)}]$ of any small subsystem $A$ fluctuates around a stationary state $\omega_A$ \cite{Gogolin2016}. That is, for most times, the local reduced density matrix $\rho_A(t)$ is close to the stationary state $\omega_A$:
\begin{align}
	D(\rho_A(t),\omega_A) \leq \delta,	
\end{align}
where $D$ denotes the trace-distance between two density matrices, defined as $D(\rho,\sigma) = \frac{1}{2}\norm{\rho-\sigma}_1$.
In general, $\omega_A$ can be any quantum state, but we say that the system \emph{thermalizes} if it is given by the reduced state of a Gibbs state: $\omega_A = \Tr_{\bar A}\left[\frac{\e^{-\beta H}}{Z_\beta}\right]$.
For generic, non-integrable (ergodic) quantum many-body systems, 
this phenomenon of \emph{thermalization} is in fact expected to occur with with precision $\delta=O(\exp(-N))$, where $N$ is the total system size \cite{Deutsch1991,Srednicki1994,Tasaki1998,Popescu2006,Rigol2008,Linden2009,Short2011,Reimann2012,Gogolin2016,Rigol2016,Gallego2017,Wilming2019,Huang2021}. This conventional view on thermalization is well established, but only takes into account the local reduced density matrix of $A$.
Recently, Refs.~\cite{Choi2021,Cotler2021,Ho2021} asked whether additional information may be obtained when, instead of tracing out the remainder of the system $\bar A$, we \emph{measure} it in an orthonormal basis $\{\ket{z}_{\bar A}\}$ and study the ensemble of post-measurement states
\begin{align}
	\ket{\psi_z} := \frac{1}{\sqrt{p_z}} (\id_A\otimes \bra{z}_{\bar A})\ket{\Psi},
\end{align}
each occuring with probability
\begin{align}
	p_z := \bra{\Psi}(\id_A\otimes\proj{z}_{\bar A})\ket{\Psi}.
\end{align}
Here and in the following, we focus on a single time $t$ and omit it from the notation.
Clearly, $\rho_A = \sum_z p_z \proj{\psi_z}$.
Interestingly, Refs.~\cite{Choi2021,Cotler2021,Ho2021}, found that for specific systems thermalizing to infinite temperature (i.e., $\omega_A=\id_A/d_A$),
the ensembles $\{(p_z,\ket{\psi_z})\}_z$ closely resemble ensembles of quantum states that are chosen uniformly at random from the set of pure states on the Hilbert space $\mc H_A$ of $A$.

In quantum information theory, such ensembles are called \emph{(approximate) quantum state designs} and, together with their generalizations of ensembles of unitary operators called \emph{(approximate) unitary designs}, are an indispensable tool in various areas of modern quantum information theory \cite{RenBluSco04,KlaRoe05,Ambainis2007,RoySco07,DankertEtAl:2009:Exact,GroAudEis07,Harrow2009,Brandao2016}, with applications ranging from benchmarking and certification of quantum devices \cite{KlieschRoth:2020:Tutorial} to the study of the black hole information paradox \cite{HaydenPreskill:2007:BlackHoles}. %TBD: Check the latter reference. A larger collection of design applications can be found in the Quantum homeopathy paper. 

Formally, an  \emph{$\epsilon$-approximate quantum state $k$-design} on a Hilbert-space $\mc H$ is defined as an ensemble of pure states $\{(p_z,\ket{\psi_z})\}_{z=1}^M$ with the property that
\begin{align}\label{eq:kdesign}
	D\Big(\,\bE_{\ket\psi\sim \mu} \proj{\psi}^{\otimes k},\ \sum\nolimits_z p_z \proj{\psi_z}^{\otimes k}\,\Big)\leq \epsilon,	
\end{align}
where $\mu$ denotes the unique normalized measure on pure quantum states induced by the Haar measure on the (special) unitary group. %This measure is unique. 
While a $1$-design simply corresponds to an ensemble whose average is identical to the maximally mixed state, higher $k$-designs also approximate higher moments of the uniform distribution on pure quantum states. 

If a system thermalizes to infinite temperature, it is therefore clear that \emph{every} measurement basis $\ket{z}_{\bar A}$ gives rise to an (approximate) 1-design. 
However, Refs.~\cite{Choi2021,Cotler2021,Ho2021} found that, for specific measurement bases, in particular the computational basis constructed from tensor-product states, the post-measurement ensemble are approximate $k$-designs for $k>1$ after sufficiently long time. 
This indicates that these post-measurement ensembles contain information that is not contained in the reduced density matrix alone and that thermalization in terms of the reduced density matrix is a weaker property than the emergence of quantum state designs as post-measurement ensembles. 
Consequently, the authors interpreted their findings as going ``beyond the conventional paradigm of quantum thermalization'' \cite{Cotler2021}.
This claim was further substantiated by the observation that even after the thermalization time, the post-measurement ensemble continued to approach quantum state designs of higher and higher precision. 

In this work, we propose a different perspective and show that whenever a small subsystem of a big quantum system is close to maximally mixed (for example, but not necessarily, due to thermalization to close to infinite temperature), then almost all measurement bases give rise to an $\epsilon$-approximate $k$-design. In other words: thermalization of the reduced density matrix alone implies the emergence of quantum state designs as post-measurement ensembles for almost all measurement bases.
Specifically, we prove the following theorem:
\begin{theorem}\label{thm:main} Consider a quantum state $\ket{\Psi}\in \mc H_A\otimes \mc H_{\bar A}$ and pick a measurement bases $\{\ket{z}_{\bar A}\}$ on $\mc H_{\bar A}$ randomly from the Haar measure. 
For $1 > \epsilon',\delta,\Delta >0$ and $k\in\mathbb N$  suppose that
	\begin{align}
		D\!\left(\Tr_{\bar A}[\,\proj{\Psi}\,],\  \frac{\id_A}{d_A}\right) \leq \delta < \frac{1}{2d_A}\,,
	\end{align}
	and $M:=\dim(\mc H_{\bar A})$ fulfills
	\begin{align}\label{eq:main1}
		M > \frac{4(2k-1)^2 d_A^{2k-1}}{{\epsilon'}^2}\log\left(\frac{2 d_A^{2k}}{\Delta}\right).
	\end{align}
	Then with probability of at least $1-\Delta$, the induced post-measurement ensemble is an $\epsilon$-approximate $k$-design with
	\begin{align}
		\epsilon \leq \epsilon' + 2k\sqrt{d_A\delta}+ d_A \delta \, .
	\end{align}
\end{theorem}
The theorem says that when $\delta,\epsilon'>0$ are some arbitrarily small constants and the system size is increased while $A$ is being held fixed, then the probability that the post-measurement ensemble is an $\epsilon$-approximate quantum state $k$-design approaches unity exponentially in $M=d_{\bar A}$. 
However, we can also let $\epsilon'$ and $\Delta$ depend on $M$. For example, if the dimension of $A$ is held fixed, we can set $\Delta = O(\exp(-\sqrt{M}))$ and $\epsilon' \leq O(M^{-1/4})$ with $\eqref{eq:main1}$ still being fulfilled. 
If we imagine that our system is a lattice of $N$ spin-1/2 systems, then $M = O(\exp(N))$, so that we find that with a probability \emph{double-exponentially close to unity} in $N$, the post-measurement ensemble is an $\epsilon$-approximate quantum state $k$-design with $\epsilon = O(\sqrt{\delta}) + O(\exp(-N))$. 
It is now important to remember that an interacting, ergodic quantum system is typically expected to thermalize with a precision $\delta = O(\exp(-N))$, where $N$ is the \emph{total} system size (see introduction). 
Therefore we can expect $\epsilon=O(\exp(-N))$ in a typical large, interacting many-body system with probability double-exponentially close to $1$ in $N$. 
However, we emphasize here, that our result does not make any assumptions about how the state $\ket{\Psi}$ arose. It is agnostic to whether we talk about a many-body system or simply a bipartite system with one small and one very large part.

In Ref.~\cite{Cotler2021} it was shown that if one samples pure states at random from the measure on pure states induced by the Haar measure, then with overwhelming probability there exists a measurement basis on $\bar A$ that induces an approximate $k$-design. 
Indeed, since the Haar measure is left- and right-invariant, this is true for any fixed measurement basis. 
As a second result, Ref.~\cite{Cotler2021} establishes that such  
measurement bases in fact already exist if the state is drawn from an approximate quantum state $k$-design. 
The technical argument of their second result is considerably more sophisticated since it requires to control higher moments using lower moments via suitable polynomial approximation. 
 
By regarding generic measurement bases, theses results now also arise as a direct consequence of our Theorem~\ref{thm:main} using only the entanglement properties of the random states:
It is well known that randomly sampling pure states gives, with overwhelming probability, a state that is locally very close to maximally mixed on $A$ \cite{Lubkin1978,Lloyd1988,Page1993}. 
Indeed, for random states one expects $\delta = O(1/M)$, which is exponentially small in the size of $\bar A$. Hence our result applies directly. 

\emph{Typical and atypical measurement bases and time-evolution. }
One may argue that in actual experiments we usually do not measure typical bases, but highly atypical ones \footnote{In fact, in real experiments we usually don't perform projective measurements at all.}. 
For example, in a spin-1/2 system we may measure the \emph{computational basis} $\ket{i_1}\ket{i_2}\cdots\ket{i_N}$, where the states $\ket{i_k}$ with $i_k=\uparrow,\downarrow$ denote the eigenstates of spin $k$ in $z$-direction. 
Indeed, Refs.~\cite{Choi2021,Ho2021} found that in concrete models, even measuring the computational basis induces approximate quantum state $k$-designs after sufficiently long time. 
Moreover the accuracy of the designs continued to increase even long after the thermalization time. 
	This also motivates the second result of Ref.~\cite{Cotler2021} mentioned above: 
If the time-evolution is modelled by a random, local quantum circuit, then the outputs are approximate quantum state $k$-designs \cite{Harrow2009,Brandao2016,Harrow2018,Haferkamp2020} and thus give rise to an approximate quantum state $k$-design as post-measurement ensemble after sufficiently long time.  
%
% The second result of Ref.~\cite{Cotler2021} implies that if time-evolution is modelled by a random, local quantum circuit, then with high probability any measurement basis gives rise to an approximate quantum state $k$-design as post-measurement ensemble after sufficiently long time.
% This is because it can be proven that the outputs of random quantum circuits are approximate quantum state $k$-designs themselves \cite{Harrow2009,Brandao2016,Harrow2018,Haferkamp2020}.
%
While our result does not yield a statement about measurements in any specific measurement basis, it shows that \emph{every} time-evolution that yields a locally close to maximally mixed state leads to an approximate quantums state $k$-design for almost all measurement bases. 

%\emph{Time-evolution.} 
Here, we gain a slightly different heuristic perspective on the emergence of increasingly more accurate quantum state designs from time evolutions even in the case where the measurement basis is given by the computational basis.   
The local entanglement entropy in generic, interacting systems is expected to grow linearly with time until it saturates at the thermalization time, which is therefore expected to be of order unity. 
Due to the at most ballistic spreading of correlations implied by Lieb-Robinson bounds \cite{Lieb1972}, after the thermalization time the system $A$ can only be strongly correlated to a part of $\bar A$ of fixed-size and hence Hilbert-space dimension. 
This implies that the number of sufficiently distinct post-measurement states is bounded by this Hilbert-space dimension. 
But a high-quality quantum state design needs increasingly many elements as its error $\epsilon$ decreases and $k$ increases.  
Therefore, larger and larger times are required. 

At the same time it is expected that the \emph{complexity} of the quantum state continues to grow for a time that is \emph{exponential in the system size} \cite{Brown2017}. 
Indeed, this can be shown rigorously for a strong notion of complexity and dynamics given by random quantum circuits \cite{Haferkamp2021}, which are often expected to capture the qualitative behaviour of chaotic Hamiltonian time-evolution (other complexity measures are directly connected to the typicality as measured by quantum state $k$-designs \cite{RiKnNick:2021:Models}). 
In other words, we expect that the time-dependent quantum state $\ket{\Psi(t)}$ continues to evolve towards a more \emph{typical} state (in the sense of the Haar measure) for very long time. 
But relative to a typical state, the computational basis on $\bar{A}$ can be expected to also act like a typical measurement basis, so that our result and that of Ref.~\cite{Cotler2021} applies with increasing precision. 

\emph{Underlying intuition.} 
While the proof of Theorem~\ref{thm:main} is somewhat involved, the basic insight that leads to it is simple. We therefore now explain this basic reasoning and then sketch the proof in the remainder of the paper. Missing details are given in the Supplemental Material. 

We first argue that the existence of a measurement basis with post-measurement ensemble constituting an approximate quantum state $k$-design is simply a question of the existence of such designs in the respective dimensions.  
To this end, imagine we are given an $\epsilon$-approximate quantum state $k$-design $\{ (p_z, \ket{\psi_z}) \}_{z=1}^M$ that is also an \emph{exact} $1$-design. 
We can inscribe the design into a pure quantum state as
\begin{align}\label{def:Phi}
	\ket{\Phi} \coloneqq \sum_{z=1}^M \sqrt{ p_z} \ket{\psi_z}\otimes\ket{z}_{\bar A},
\end{align}
where $\{\ket{z}_{\bar A}\}$ is some orthonormal basis on $\bar A$. 
The design properties manifest itself in $\ket\Phi$ as follows:
\begin{enumerate}[i)]
	\item\label{1design} The state $\ket{\Phi}$ is maximally mixed reduced to $A$: 
		\begin{align}\Tr_{\bar A}[\proj{\Phi}]=\sum_z p_z\proj{\psi_z} = \id_A/d_A.\end{align}
		\item\label{purifiedkdesign} Measuring $\ket{\Phi}$ on $\bar A$ in the basis $\ket{z}_{\bar A}$ induces the approximate quantum state $k$-design $\{(p_z,\ket{\psi_z})\}_{z=1}^M$ on $A$ as post-measurement ensemble. 
\end{enumerate}
For simplicity consider now some quantum state $\ket{\Psi}$ that is \emph{exactly} thermalized to inifinite temperature. 
Then both $\ket\Psi$ and $\ket\Phi$ are purifications of a maximally mixed state on $A$ with the same purifying Hilbert-space $\mc H_{\bar A}$. 
Hence, there exist a unitary $U$ acting on the purifying space such that 
\begin{align}
	\ket\Psi = \mb 1_A\otimes U\ket\Phi.
\end{align}
Therefore, by \ref{purifiedkdesign}), measuring $\ket{\Psi}$ in the orthonormal basis defined by $U\ket{z}$ yields the $\epsilon$-approximate quantum state $k$-design $\{(p_z,\psi_z)\}_{z=1}^M$ as post-measurement states. 
What needs to be proved to obtain Theorem~\ref{thm:main} is therefore that a) the above reasoning carries over to the approximate case and b) appropriate quantum state designs exists and, in fact, arise with high probability from a random measurement basis. We now explain these parts separately and then combine them into the proof of Theorem~\ref{thm:main}.

\emph{Approximate thermalization. } 
The basic object of our study is the \emph{moment operator} 
\begin{align}
	M^{(k)}_{\ket \Psi} = \sum_z p_z \proj{\psi_z}^{\otimes k}
\end{align}	
associated to the post-measurement ensemble $\{(p_z,\ket{\psi_z})\}_{z=1}^M$ obtained from $\ket{\Psi}$.
We want to show that it is close to $M^{(k)}_{\mathrm{Haar}}=\mathbb E_{\psi\sim\mu} \proj{\psi}^{\otimes k}$, where $\mu$ is the Haar measure. 
Our first technical result shows that the moment operator only weakly depends on the reduced density matrix $\rho_A$ as long as it is close to maximally mixed.
This allows us to reduce the general analysis to the case of perfect thermalization for the price of a small additive error.
The result is summarized in the following lemma, whose proof is given in the Supplemental Material. 
\begin{lemma}[Continuity]\label{lemma:continuity}
	Let $\ket{\Psi} = \sum_i \sqrt{q_i} \ket{i}_A\ket{\phi_i}_{\bar A}$ and $\ket{\Phi} = \frac{1}{\sqrt{d_A}}\sum_i \ket{i}_A\ket{\phi_i}_{\bar A}$ be Schmidt-decompositions of two pure states.  
	Moreover, assume that the reduced state $\rho_A=\sum_i q_i\proj{i}_A$ of $\ket\Psi$ fulfils $D(\rho_A,\id_A/d_A)\leq \delta$ with $\delta\leq 1/(2d_A)$. Then
	\begin{align}
		D(M^{(k)}_{\ket \Psi},M^{(k)}_{\ket \Phi}) \leq 2k\sqrt{d_A\delta}+\delta d_A.
	\end{align}
\end{lemma}
The Lemma says that whenever $\rho_A$ is sufficiently mixed, instead of $\ket\Psi$ we can equivalently look at the associated state $\ket{\Phi}$ that is locally perfectly thermalized.

\emph{The case of perfect thermalization for general measurement bases.}  
It remains to establish the existence of the quantum state $k$-design that we want to encode. 
Deterministic construction of quantum state $k$-designs (in particular, tight ones) is a well-studied task in math and quantum information related to important problems, see e.g.~Ref.~\cite{BengtssonZyczkowski:2017:Geometry:Designs}.  
For our purpose a simpler probabilistic construction suffices.
The guiding example is the following: 
A number of $M$ states $\ket{\psi_z}$ drawn independently from an exact quantum state $k$-design constitute an $\epsilon$-approximate quantum state $k$-design with respect to the uniform measure with $\epsilon = O(\exp(-M))$. 
(This can be quicky seen by standard matrix concentration results, see also Ref.~\cite{AmbainisEtAl:2009} for the analogous statement for unitary designs.) 
Drawing the $\ket{\psi_z}$ independently at random does, however, not yield an exact $1$-design corresponding to the requirement (i) of exact thermalization.
To understand this additional constraint, consider the linear operator from $\mc H_A \simeq \mathbb C^{d_A}$ to $\mc H_{\bar A}\simeq \mathbb C^M$ defined as 
\begin{align}
	W = \sqrt{d_A} \sum_z \sqrt{p_z} \ketbra{z}{\psi_z},
\end{align}
which is in one-to-one correspondence with the state $\ket\Phi = \sum_z \sqrt{p_z} \ket{\psi_z}\otimes \ket{z}$.
{Instead of changing the measurement basis from $\ket{z}_{\bar A}$ to $U\ket{z}_{\bar A}$ with a unitary $U$, 
we can equivalently change the state $\ket\Phi$ to $(\id\otimes U^\dagger)\ket\Phi$ and keep the measurement basis fixed. 
From now on we will therefore describe a change of the measurement basis by actively transforming $\ket\Phi$ accordingly.
The associated isometry then transforms as $W\mapsto V(U) := U^\dagger W$.
}
In the following, we will often omit the argument $U$ from $V$.

The requirement (i) now naturally translates into $V$ being an \emph{isometry} between the Hilbert-spaces $\mc H_A$ and $\mc H_{\bar A}$, i.e. $V^\dagger V = \id_A$.
Observe that, up to normalization, the row-vectors $\bra{\psi_z}$ of the post-measurement ensemble are simply given by the rows of the isometry $V$.
Choosing a random such isometry from the measure on the set of isometries that is left-invariant under multiplication by unitaries corresponds to a Haar random choice of measurement basis.
It can be done by choosing a Haar-random unitary $U$ and constructing $V(U)=UW$, which is independent of $W$ as a random variable.

This discussion motivates to probabilistically construct an approximate quantum state $k$-design but \emph{exact} $1$-design as follows: 
\emph{Draw a random isometry $V=\sum_z \ketbra{z}{v_z}$ with row vectors $\bra{v_z}$ in some basis. 
Choose as vectors of the design the vectors $\ket{\psi_z} = \ket{v_z} / {\|\ket{v_z}\|}$ and associate the probability mass 
$p_z=\|\ket{\hat v_z}\|^2 / d_A$. }
We will call the ensemble $(p_z,\psi_z)_{z=1}^M$ obtained from this procedure the \emph{row-ensemble induced by $V$}.

To summarize, given a measurement basis, we have expressed the post-measurement states (up to normalization) as rows of an isometry $V$ uniquely associated to $\ket\Phi$. 
The question of whether the post-measurement ensemble obtained from $\ket{\Phi}$ is an approximate quantum state $k$-design is equivalent to asking whether the row-ensemble induced by the isometry $V$ is an approximate quantum state $k$-design. 
Taking the measurement basis at random amounts to selecting $V$ at random from the unitarily invariant measure on the isometries. 

By construction the row ensemble induced by an isometry $V$ is an exact $1$-design. 
At the same time, using the unitarily invariant measure on the isometries, the construction is very similar to the guiding example where the states are drawn completely independent from each other. 
The second technical ingredient to our main result is to formally prove that the row ensemble induced by a random isometry $V: \mathbb C^{d_A}\rightarrow \mathbb C^{M}$ forms an approximate quantum state $k$-design on $\mathbb C^{d_A}$ if $M$ is sufficiently large. 
This is summarized in the following lemma:

\begin{lemma}[Random isometries induce approximate quantum state designs]\label{lemma:randomisometries}
	Let $V:\mathbb C^{d_A}\rightarrow \mathbb C^M$ be a random isometry sampled from the unitarily invariant measure and $\{\ket z\}_{z=1}^M$ an orthonormal basis on $\mathbb C^M$. Then if $1>\epsilon,\Delta>0$ and 
	\begin{align}
		M > \frac{4(2k-1)^2 d_A^{2k-1}}{\epsilon^2}\log\left(\frac{2d_A^{2k}}{\Delta}\right)
	\end{align}
	the row ensemble induced by $V$ is an $\epsilon$-approximate quantum state $k$-design with probability at least $1-\Delta$. 
\end{lemma}
This result on a probabilistic construction of approximate quantum state designs is of independent interest from a quantum information perspective and might find other applications. 
The proof of the lemma can be found in the Supplemental Material. 
It is based on two essential ingredients: First, the unitary group on $\mathbb C^{d_A}$ embeds into the unitary group on $\mathbb C^M$ via an embedding $U_A\mapsto \hat U_A$ in such a way {that $V(\hat U_A U) =V(U)U_A^\dagger$.} 
This can be used to show that the expected moment operator of the induced ensemble matches $M^{(k)}_{\mathrm{Haar}}$.
Second, we show that the map from $U$ to the associated moment operator of the induced ensemble is Lipschitz-continuous. We can therefore use the phenomenon of \emph{concentration of measure} on the special unitary group and the Lévy-Gromov Lemma \cite{Milman1986,Ledoux2001,Anderson2009} to prove that its extremely unlikely that the moment operator significantly deviates from the expected value.   

\emph{Proof of Theorem~\ref{thm:main}. }We are now in position to prove Theorem~\ref{thm:main}: Let $\ket{\Psi}$ be a state with reduced density matrix $\rho_A$ fulfiling $D(\rho_A,\id_A/d_A)\leq \delta < 1/(2d_A)$ as before and suppose we measure it in the basis $U\ket{z}_{\bar A}$, where $U$ is a random unitary. 
Then, by Lemma~\ref{lemma:continuity}, 
the moment operator of {the associated} exactly thermalized state $\ket\Phi$ deviates only as $D(M_{\ket\Psi}^{(k)},M_{\ket\Phi}^{(k)}) \leq 2k\sqrt{d_A\delta}+\delta d_A$. 
The post-measurement ensemble of $\ket\Phi$, when measured in the basis defined by $U$, is the row ensemble induced by a random isometry. 
Hence, if $M$ fulfills the conditions of the theorem, then Lemma~\ref{lemma:randomisometries}
shows that 
with probability at least $1-\Delta$ we have $D(M_{\ket\Phi}^{(k)},M_{\mathrm{Haar}}^{(k)})\leq \epsilon'$. 
Combining both bounds, we conclude that with probability at least $1-\Delta$ it holds that $D(M_{\ket\Psi}^{(k)},M_{\mathrm{Haar}}^{(k)})\leq \epsilon' + 2k\sqrt{d_A\delta} +\delta d_A$, which proves the theorem.

\emph{Conclusions. }
We have shown that when any sufficiently high-dimensional purification of an (approximately) maximally mixed state is measured in a \emph{random basis} on the purifying space, then it yields an ensemble of quantum states that is almost uniformly distributed.
This is true for any bipartite system in a pure state, where a small subsystem is close to maximally mixed, regardless of how the state arose. 
In particular, this is the case when a large, complex quantum system locally thermalizes close to infinite temperature. 
This shows that the emergence of quantum state designs reported in Refs.~\cite{Choi2021,Cotler2021,Ho2021} is generic for such thermalizing systems if we consider general measurement bases. 
Interesting open problems that remain are: Can one show that this emergence of quantum state designs is true for generic, local Hamiltonians in the long-time limit even for atypical measurement bases such as the computational basis? 
Can one use Lieb-Robinson bounds or similar arguments to provide a lower-bound on the time required for the computational basis measurements to induce an approximate quantum state $k$-design? 
A related technical task of independent interest, is the partial de-randomization of our construction by replacing Haar random measurement basis with those obtained from a unitary $k$-designs. 
Finally, it would be interesting to study the properties of post-measurement ensembles induced from random measurement bases on states that are locally thermal with a finite (instead of close to infinite) temperature.

\begin{acknowledgments}
H.W. would like to thank Joe Renes for discussions in an early stage of this project. 
I.R.\ would like to thank Jonas Haferkamp, Micha\l\ Oszmaniec and Zolt\'an Zimbor\'as for discussions. 
Support by the DFG through SFB 1227 (DQ-mat), Quantum Valley Lower Saxony, and funding
 by the Deutsche Forschungsgemeinschaft (DFG, German Research Foundation) under Germanys Excellence Strategy EXC-2123 QuantumFrontiers 390837967 is also
	acknowledged. 
\end{acknowledgments}
\newpage
\bibliographystyle{apsrev4-1}
\bibliography{bibliography_designs}
\clearpage
\appendix
\onecolumngrid
\section{Auxiliary Lemmas}
In this section we collect two auxiliary Lemmas that we require for our proofs. 
For vectors $\vec p, \vec r \in \CC^d$, let $D(\vec p, \vec r) \coloneqq \frac12 \|\vec p - \vec r\|_{\ell_1}$.
\begin{lemma}[Trace-distance of mixtures]\label{lemma:mixtures}
	Let $\{(p_z,\rho_z)\}_z$ and $\{(r_z,\sigma_z)\}$ be ensembles of density matrices. Then
	\begin{align}
		D\big(\sum_z p_z \rho_z^{\otimes k},\sum_z r_z \sigma_z^{\otimes k}\big) &\leq \sum_z p_z D(\rho_z^{\otimes k},\sigma_z^{\otimes k}) + D(\vec p,\vec r) \leq k \sum_z p_z D(\rho_z,\sigma_z) + D(\vec p,\vec r).
	\end{align}
	\begin{proof}
		The first inequality follows from the triangle-inequality. Excplicitly:
		\begin{align}
			\norm{\sum_z p_z \rho_z^{\otimes k} - \sum_z r_z \sigma_z^{\otimes k}}_1 &\leq \norm{\sum_z p_z \left(\rho_z^{\otimes k}-\sigma_z^{\otimes k}\right)}_1 \nonumber + \norm{\sum_z (p_z-r_z)  \sigma_z^{\otimes k}}_1  \\
			&\leq \sum_z p_z \norm{\rho_z^{\otimes k}-\sigma_z^{\otimes k}}_1 + \sum_z |p_z-r_z|\\
			&=2\sum_z p_z D(\rho_z^{\otimes k},\sigma_z^{\otimes k}) + 2D(\vec p,\vec r).
		\end{align}
		The second inequality follows from the submultiplicativity of the Schatten norms using a telescoping sum ($\norm{\rho_z}_1=\norm{\sigma_z}_1=1$):
		\begin{align}
			\norm{\rho_z^{\otimes k} - \sigma_z^{\otimes k}}_1 &= \norm{\rho_z^{\otimes k} - \rho_z\otimes\sigma_z^{\otimes k-1}}_1+ \norm{\rho_z-\sigma_z}_1\norm{\sigma_z^{\otimes k- 1}}_1 \\
			&\leq  \norm{\rho_z^{\otimes k-1} - \sigma_z^{\otimes k-1}}_1 + \norm{\rho_z-\sigma_z}_1 \\
			&\leq \ldots \leq k\norm{\rho_z-\sigma_z}_1.
		\end{align}
	\end{proof}
\end{lemma}

\begin{lemma}[Derivative]\label{lemma:derivative} Let $\rho(\lambda)_z$ for $z=1,\ldots,M$ denote differentiable curves of positive semidefinite operators with $p(\lambda)_z:=\Tr[\rho(\lambda)_z]>0$. Define $\hat\rho(\lambda)_z = \rho(\lambda)_z/p(\lambda)_z$. Then
	\begin{align}
		\frac{\rmd}{\rmd \lambda} \sum_z p(\lambda)_z \hat\rho(\lambda)_j^{\otimes k} = \sum_{l=0}^{k-1} \sum_z  \hat \rho(\lambda)_z^{\otimes l} \otimes \rho'(\lambda)_z\otimes \hat\rho(\lambda)_z^{k-(l+1)} - (k-1) \sum_z p'(\lambda)_z \rho(\lambda)_z^{\otimes k}. 
	\end{align}
	\begin{proof}
	Direct calculation yields:
		\begin{align}
			\frac{\rmd}{\rmd \lambda} \sum_j p(\lambda)_j \hat\rho(\lambda)_j^{\otimes k} &= \sum_z p'(\lambda)_z \hat\rho(z)^{\otimes k}_z + \sum_{l=0}^{k-1}\sum_z p(\lambda)_z \hat\rho(\lambda)_z^{\otimes l} \otimes \left[\frac{\rho'(\lambda)}{p(\lambda)_z} -\frac{p'(\lambda)_z}{p(\lambda)_z}\hat \rho(\lambda)_z \right]  \otimes \hat\rho(\lambda)_z^{\otimes k - (l+1)}\\
			&= \sum_{l=0}^{k-1} \sum_z \hat\rho(\lambda)_z^{\otimes l} \otimes \rho'(\lambda)_z \otimes \hat\rho(\lambda)_z^{\otimes k-(l+1)} - (k-1) \sum_z p'(\lambda)_z \hat\rho(\lambda)_z^{\otimes k}.
		\end{align}
	\end{proof}
\end{lemma}

%IR: Fuchs--van de Graaf inequalities are a bit overkill for pure states since one can easily calculate the exact expression there directly. 
% \begin{lemma}[Fuchs--van de Graaf inequalities \cite{Fuchs1999}]\label{lemma:fuchs} Fidelity $F$ and trace-distance $D$ are related as
% 	\begin{align}
% 	1-F(\rho,\sigma) \leq  D(\rho,\sigma) \leq \sqrt{1-F(\rho,\sigma)^2}.
% 	\end{align}
% 	Moreover, if one of the states is pure, we have 
% 	\begin{align}
% 	1-F(\rho,\sigma)^2 \leq D(\rho,\sigma).
% 	\end{align}
% \end{lemma}
\section{Proof of Lemma~\ref{lemma:continuity}}
Here we prove Lemma~\ref{lemma:continuity} from the main text. Recall that we consider the states 
\begin{align}
	\ket\Psi &= \sum_i \sqrt{q_i} \ket{i}_A\otimes\ket{\phi_i}_{\bar A},\\
	\ket\Phi &= \frac{1}{\sqrt{d_A}} \sum_i \ket{i}_A\otimes\ket{\phi_i}_{\bar A}, 
\end{align}
where $\{\ket{i}_A\}_{i=1}^{d_A}$ is an orthonormal basis on $\mc H_A$ and the vectors $\ket{\phi_i}_{\bar A}$ are part of an orthonormal basis on $\mc H_{\bar A}$.

Furthermore, the reduced state $\rho_A = \sum_i q_i \proj{i}_A$ of $\ket\psi$ fulfills (by assumption) $D(\rho_A,\id_A/d_A)\leq \delta$ with $\delta<1/(2d_A)$. 
We denote the post-measurement states as $\psi_z = \proj{\psi_z}$ and $\varphi_z = \proj{\varphi_z}$ occuring with probabilities $p_z$ and $r_z$, respectively, so that
\begin{align}
	M^{(k)}_{\ket\Psi} = \sum_z p_z \psi_z^{\otimes k},\quad M^{(k)}_{\ket\Phi} = \sum_z r_z \varphi_z^{\otimes k}.
\end{align}
We wish to show that
\begin{align}
	D(M^{(k)}_{\ket\Psi},M^{(k)}_{\ket\Phi}) \leq 2k\sqrt{d_A\delta} + \delta.
\end{align}
As in the main-text, we introduce a suitable isometry given by
\begin{align}
	V = \sqrt{d_A}\sum_z \sqrt{r_z} \ketbra{z}{\varphi_z}.
\end{align}
Then, since
\begin{align}
	\ket\Psi = (\sqrt{d_A \rho_A}\otimes\id)\ket\Phi,
\end{align}
the post-measurement states may be written using the isometry $V$ as
	\begin{align}
		\ket{\psi_z} = \frac{\sqrt{\rho_A}V^\dagger\ket{z}}{\sqrt{\bra{z} V\rho_AV^\dagger \ket{z}}},\quad \ket{\varphi_z} = \frac{V^\dagger\ket{z}}{\sqrt{\bra{z} VV^\dagger\ket{z}}},
	\end{align}
and $p_z = \bra{z}V\rho_AV^\dagger\ket{z},r_z = \frac{1}{d_A} \bra{z} VV^\dagger\ket{z}$.

Observing that the moment operators are convex mixtures of tensor products of pure states, we first make use of the Lemma~\ref{lemma:mixtures} to get
\begin{align}
	D(M^{(k)}_{\ket\Psi},M^{(k)}_{\ket\Phi}) \leq k \sum_z p_z D(\psi_z,\varphi_z) + D(\vec p,\vec r).
\end{align}
We first deal with the second term using the following Lemma:
	\begin{lemma}\label{lemma:normalization}
	Under the above assumptions, we have $|p_z-r_z| \leq r_z 2 \delta d_A$ and hence $D(\vec p,\vec r)\leq \delta d_A$.
	\begin{proof}
		We have $r_z d_A = \norm{V^\dagger \ket{z}}^2 = \norm{V^\dagger\proj{z}V}$. Hence
		\begin{align}
			|p_z-r_z| &= \left|\sandwich{z}
			{V\left(\rho_A - \frac{\id_A}{d_A}\right)V^\dagger}{z}\right| = \left|\Tr\left[\left(\rho_A-\frac{\id_A}{d_A}\right)V^\dagger\proj{z}V\right]\right|\\
			&\leq \norm{\rho_A-\frac{\id_A}{d_A}}_1 \norm{V^\dagger\proj{z}V} = 2D\!\left(\rho_A,\frac{\id_A}{d_A}\right)r_z d_A \leq 2\delta r_z d_A,
		\end{align}
		where we used the matrix Hölder inequality $|\Tr[A^\dagger B]|\leq \norm{A}_1 \norm{B}$ and the definition of the trace-distance together with $\norm{\proj{\psi}}=\norm{\ket{\psi}}^2$. 
	\end{proof}
\end{lemma}
We now turn to the first term via the following lemma. 
\begin{lemma} Under the same assumption as above and, in particular, $\delta \leq 1/(2d_A)$, we have
	\begin{align}			
		D(\psi_z,\varphi_z) \leq 2\sqrt{d_A\delta}. 
	\end{align}
	\begin{proof}
		Since the states are pure, it is straight-forward to calculate that 
		\begin{align}\label{eq:fuchs}
			D(\psi_z, \phi_z) = \sqrt{1 - \Tr[\psi_z\phi_z]}\,
		\end{align}
		see e.g.\ \cite[Lemma 14]{KlieschRoth:2020:Tutorial}. 
		(This is case where the upper bound of the Fuchs-van de Graf inequalitites \cite{Fuchs1999} relating trace distance and fidelity are tight.) 
		% First, since the states are pure we have by the Fuchs-van de Graf inequalitites (Lemma~\ref{lemma:fuchs})
		% \begin{align}\label{eq:fuchs}
		% 	D(\psi_z,\varphi_z) =	\sqrt{1 - F(\psi_z,\varphi_z)^2}.
		% \end{align}
		Writing out the overlap we have
		\begin{align}
			\sqrt{\Tr[\psi_z\phi_z]} &= \frac{1}{\sqrt{r_z p_z d_A}}\bra{z}V\sqrt{\rho_A}V^\dagger\ket{z}.
		\end{align}
		We first bound the prefactor using Lemma~\ref{lemma:normalization} as
		\begin{align}
			\frac{1}{\sqrt{r_z p_zd_A}}\geq \frac{1}{\sqrt{r_z^2d_A(1+2\delta)d_A}} \geq \frac{1}{r_z \sqrt{d_A(1+2\delta)d_A}}.
%			\frac{M}\sqrt{1 + M \Delta_}} \geq M\left(1- M\Delta_{\max}\right).
		\end{align}
		For the remainder, we use the assumption $D(\rho_A,\id_A/d_A)\leq \delta \leq 1/(2d_A)$ to get
		\begin{align}
			q_i \geq \frac{1}{d_A} - 2\delta = \frac{1}{d_A}(1-2d_A\delta).
		\end{align}
		Therefore $\sqrt{\rho_A} \geq \frac{1}{\sqrt{d_A}}\sqrt{1-2d_A\delta}\id_A$ and
		\begin{align}
			\bra{z}V \sqrt{\rho_A}V^\dagger\ket{z} \geq \frac{d_A r_z}{\sqrt{d_A}}\sqrt{1-2d_A\delta} = r_z \sqrt{d_A(1-2d_A\delta)}.
		\end{align}
		Putting both bounds together, we arrive at 
		\begin{align}
			\sqrt{\Tr[\psi_z\phi_z]} \geq \sqrt{\frac{1-2d_A\delta}{1+2\delta d_A}}
		\end{align}
		We thus find
		\begin{align}
			D(\psi_z,\varphi_z) \leq \sqrt{\frac{2(d_A+1)\delta}{1+2\delta d_A}} \leq 2\sqrt{d_A\delta } 
		\end{align}
	\end{proof}
\end{lemma}
All in all, we have shown 
\begin{align}
	D(M^{(k)}_{\ket\Psi},M^{(k)}_{\ket\Phi}) \leq \delta d_A + 2k\sqrt{d_A\delta}  .
\end{align}

\section{Proof of Lemma~\ref{lemma:randomisometries}}
Here we provide the proof of Lemma~\ref{lemma:randomisometries}. Since the isometry $V$ depends on the random variable $U$, we will write $V(U)$ in the following:
\begin{align}
	V = V(U)= U W,
\end{align}
where $W$ is an isometry from $\mathbb C^{d_A}$ to $\mathbb C^M$.
Note that here, in contrast to the main text, we have omitted {Hermitian} conjugation from the unitary in the definition of $V(U)$. 
This is insignificant since $U$ is a random unitary, but lightens the notation in the following.
We now also view the moment operator as a function of the random unitary $U$ and therefore write
\begin{align}
	M^{(k)}_U &\coloneqq \sum_z p_z(U) R_z(U)^{\otimes k}, \\
	R_z(U) &=\frac{V(U)^\dagger \proj{z} V(U)}{d_A p_z(U)},\\
	p_z(U) &=\bra{z}V(U)\frac{\id_A}{d_A} V(U)^\dagger\ket{z}.
\end{align}
We will now first show that $M^{(k)}_U$ is equal to $M^{(k)}_{\mathrm{Haar}}$ in expectation and then show that large deviations from the expected value are extremely unlikely.
In the following it will be important that the unitary group $\mathrm{SU}(d_A)$ embeds into $\mathrm{SU}(M)$ in the following way:
{Let $\ket{i}_A$ be the standard basis on $\mathbb C^{d_A}$. We define the vectors $\ket{\chi_i} := W\ket{i}_A$  for $i=1,\ldots,d_A$ and extend them to an orthonormal basis $\ket{\chi_z}$ on $\mathbb C^M$ with $z=1,\ldots,M$. 
Then  $\bra{\chi_i}W\ket{j} = \delta_{ij}$.
In other words, with this choice of bases we can write $\mathbb C^M = \mathbb C^{d_A}\oplus\mathbb C^{M-d_A}$ and the mapping $W$ simply takes the form 
\begin{align}
	W = \begin{pmatrix} \id_{d_A} \\ 0_{(M-d_A)\times d_A} \end{pmatrix} = \id_{d_A} \oplus 0_{(M-d_A)\times d_A},
\end{align}
where $0_{m\times n}$ is the zero $m\times n$-matrix. Given an element $U_A\in \mathrm{SU}(d_A)$ with matrix elements $(U_A)_{ij} = \bra{i}U_A\ket{j}$, we then define $\hat U_A\in \mathrm{SU}(M)$ in the basis $\{\ket{\chi_z}\}_{z=1}^M$ as
$\hat U_A = U_A \oplus \id_{M-d_A}$, 
so that $\bra{\chi_i} \hat U_A \ket{\chi_j} = \bra{i}U_A\ket{j}$ for $i,j\leq d_A$ and $\bra{\chi_z}\hat U_A\ket{\chi_{z'}}=\delta_{z,z'}$ otherwise.
With these definitions, we have
\begin{align}\label{eq:covariance}
	V(U \hat U_A) = U (U_A\oplus \id)(\id \oplus 0) = U (\id\oplus 0)U_A = UWU_A = V(U)U_A,
\end{align}	
where we dropped subscripts for better readability.}
Since $V(U)V(U)^\dagger$ is an orthonormal projection with rank $d_A$, $\mathbb E_U p_z(U) = 1/M$, where $\mathbb E_U$ is a shorthand for $\mathbb E_{U\sim \mu(\mathrm{SU}(M))}$.
Using the right-invariance of the Haar measure and \eqref{eq:covariance}, we can now compute the expected moment operator as
\begin{equation}
\begin{split}
	\mathbb E_U M_U^{(k)} &= \sum_z \mathbb E_U p_z(U) R_z(U)^{\otimes k} \\
	&= \sum_z \mathbb E_U \mathbb E_{U_A} p_z(U \hat U_A) R_z(U \hat U_A)^{\otimes k} \\
	%&= \sum_z \mathbb E_U  p_z(U) \mathbb E_{U_A}R_z(U U_A)^{\otimes k} \\
	&= \sum_z \mathbb E_U p_z(U) \mathbb E_{U_A} (U_A^\dagger R_z(U) U_A)^{\otimes k}\\
	&= \frac{1}{M} \sum_z M^{(k)}_{\mathrm{Haar}} = M^{(k)}_{\mathrm{Haar}},
\end{split}
\end{equation}
where it is understood that $U_A$ is sampled from the Haar measure on $\mathrm{SU}(d_A)\subseteq \mathrm{SU}(M)$. To arrive at the last line, we used that $R_z(U)$ is a pure state, so that $\mathbb E_{U_A} (U_A^\dagger R_z(U) U_A)^{\otimes k}=M^{(k)}_{\mathrm{Haar}}$. 
Thus, in expectation the post-measurment ensemble coincides with a Haar random ensemble of pure states.

What is left to show is that with overwhelming probability $M_U^{(k)}$ is also close to $M^{(k)}_{\mathrm{Haar}}$ for individual samples of $U$. 
To do so, we make use of the fact that the group $\mathrm{SU}(M)$ shows the phenomenon of \emph{concentration of measure}, which is summarized in the following Lemma. 
It says that for any differentiable function on $\mathrm{SU}(M)$, significant deviations from its mean are extremely unlikely if $M$ is large. To state the Lemma, we define for any real-valued differentiable function $f$ on $\mathrm{SU}(M)$ the constant $L(f)^2 := \sup_{U} \Tr[\nabla f(U)^\dagger \nabla f(U)]$.
\begin{lemma}[L\'evy-Gromov \cite{Milman1986,Ledoux2001,Anderson2009}] Let $f:\mathrm{SU}(M)\rightarrow \mathbb R$ be a differentiable function. Consider $U\in\mathrm{SU}(M)$ as random variable sampled from the Haar measure. Then
	\begin{align}
		\mathrm{Prob}\left[|f(U) - \mathbb E_U f(U)| \geq \epsilon \right] \leq 2 \exp\left(-\frac{M}{4}\frac{\epsilon^2}{L(f)^2}\right).\nonumber
	\end{align}
\end{lemma}
We will apply the L\'evy-Gromov Lemma to the functions $f_\alpha(U):=\Tr[M_U^{(k)} X_\alpha]$, where the $d_A^{2k}$ operators $X_\alpha$ form a hermitian operator basis such that $\Tr[X_\alpha^\dagger X_\beta]=\delta_{\alpha,\beta}$. We then have (see section~\ref{sec:lipschitz} for details)
\begin{align}\label{eq:bound1}
	\mathrm{Prob}\left[ D( M_U^{(k)},M^{(k)}_{\mathrm{Haar}})>\epsilon'\right] &\leq \sum_\alpha \mathrm{Prob}\left[|f_\alpha(U)-\mathbb E_U f_\alpha(U)| > \frac{2\epsilon'}{d_A^k}\right],
\end{align}
which follows from monotonicity of norms and a union bound. 
The main technical contribution that is therefore left is to show that the constants $L(f_\alpha)$ are small. In section~\ref{sec:lipschitz} we show $L(f_\alpha)\leq 2(2k-1)/\sqrt{d_A}$. 
Inserting this bound into \eqref{eq:bound1} then yields
\begin{align}\label{eq:boundfinal}
	\mathrm{Prob}&\left[ D( M_U^{(k)},M^{(k)}_{\mathrm{Haar}})>\epsilon'\right] \leq 2d_A^{2k}  \exp\left(-\frac{M}{4(2k-1)^2 d_A^{2k-1}}{\epsilon'}^2\right).
\end{align}

\section{Applying the L\'evy-Gromov Lemma and bounding the constants $L(f_\alpha)$}
\label{sec:lipschitz}
In this section we provide the details on how to arrive at the final bound \eqref{eq:boundfinal}.
We consider the functions $f_\alpha:\mathrm{SU}(N)\rightarrow\mathbb R$ defined by
\begin{align}
	f_\alpha(U) = \Tr[M_U^{(k)} X_\alpha],
\end{align}
where the $d_A^{2k}$ operators $X_\alpha$ form an orthonormal (wrt. Hilbert-Schmidt inner product), hermitian operator-basis on $\mc H_A^{\otimes k}$.
We have
\begin{align}
	M_U^{(k)} = \sum_\alpha f_\alpha(U) X_\alpha
\end{align}
and
\begin{align}
	M_{\mathrm{Haar}}^{(k)} = \sum_\alpha \mathbb E_U f_\alpha(U) X_\alpha.
\end{align}
We can therefore express the Schatten-$2$-norm difference of the moment operator to the average as
\begin{align}\label{eq:two-norm-falpha}
	\norm{M_U^{(k)} - M_{\mathrm{Haar}}^{(k)}}_2 = \sqrt{\sum_{\alpha=1}^{d_A^{2k}} (f_\alpha(U) - \mathbb E_U f_\alpha(U))^2}.
\end{align}
Our strategy now is to use the L\'evy-Gromov Lemma for each function separately and then employ a union bound 
to estimate the probability that at least one of them deviates significantly from its mean. 
In detail, we first use the fact that on a $d$-dimensional Hilbert-space, we have $\norm{\cdot}_1 \leq \sqrt{d}\norm{\cdot}_2$ and then \eqref{eq:two-norm-falpha} to get
\begin{align}
	\mathrm{Prob}\left[\norm{M_U^{(k)} - M_{\mathrm{Haar}}^{(k)}}_1 > \delta \right]  &\leq \mathrm{Prob}\left[\norm{M_U^{(k)} - M_{\mathrm{Haar}}^{(k)}}_2 > \frac{\delta}{d_A^{k/2}} \right] \notag\\
	&\leq \mathrm{Prob}\left[\exists \alpha:\ |f_\alpha(U) - \mathbb E_U f_\alpha(U)|> \frac{\delta}{d_A^k}\right].	
\end{align}
We now employ the union bound and the L\'evy-Gromov Lemma to get
\begin{align}
	\mathrm{Prob}\left[\exists \alpha:\ |f_\alpha(U) - \mathbb E_U f_\alpha(U)|> \frac{\delta}{d_A^k}\right] &\leq \sum_\alpha \mathrm{Prob}\left[|f_\alpha(U) - \mathbb E_U f_\alpha(U)|> \frac{\delta}{d_A^k}\right] \notag\\
	&\leq 2d_A^{2k}\exp\left(-\frac{M\delta^2}{4 L^2 d_A^{2k}}\right),
\end{align}
where $L=\max_{\alpha}L(f_\alpha)$. It remains to bound $L$.

%Since the functions $f_\alpha$ are differentiable in $U$ (since $M_U^{(k)}$ is), we can bound the Lipschitz constant $L$ using the norm of the gradient of $f_\alpha$.
Recall that the gradient of $f_\alpha$ is defined as the (matrix-valued) function on $\mathrm{SU}(M)$ that for any curve $U_\lambda$ parametrized by $\lambda$ fulfills
\begin{align}
	\Tr[\nabla f_\alpha(U_\lambda)^\dagger U'_\lambda]= \frac{\rmd}{\rmd\lambda }f_\alpha(U_\lambda) =: f'_\alpha(U_\lambda),
\end{align}
where $U'_\lambda = \rmd/\rmd \lambda\, U_\lambda$.
We then have that
\begin{align}\label{eq:lconstant}
	L(f_\alpha) = \sup_U \sqrt{\Tr[\nabla f_\alpha(U)^\dagger \nabla f_\alpha(U)]} = \sup_U \norm{\nabla f_\alpha(U)}_2\,.
\end{align}
Using Lemma~\ref{lemma:derivative} with $p(\lambda)_z = p_z(U_\lambda)$ and $\rho(\lambda)_z = p_z(U_\lambda)R_z(U_\lambda)$ yields 
\begin{align}\notag
	f'_\alpha(U_\lambda) &= \sum_{l=0}^k \sum_z \Tr[X_\alpha R_z(U_\lambda)^{\otimes l}\otimes \left(\frac{\rmd}{\rmd \lambda} \frac{W^\dagger U_\lambda^\dagger \proj{z} U_\lambda W}{{d_A}}\right)\otimes R_z(U_\lambda)^{k-(l+1)}] - (k-1) \sum_z \Tr[X_\alpha R_z(U_\lambda)^{\otimes k}] p'_z(U_\lambda) \\
	&= \sum_{l=0}^k \sum_z \Tr[A_{\alpha,l,z}(\lambda)^\dagger U'_\lambda]  - (k-1)\sum_z \Tr[X_\alpha R_z(U_\lambda)^{\otimes k}] \Tr[\nabla p_z(U_\lambda)^\dagger U'_\lambda],
\end{align}
	where we implicitly defined the operators $A_{\alpha,l,z}(\lambda)=A_{\alpha,l,z}(U_\lambda)$, which we will evaluate later. Hence
\begin{align}
	\nabla f_\alpha(U) = \sum_z \left( \sum_{l=0}^k A_{\alpha,l,z}(U) - (k-1) \Tr[X_\alpha R_z(U)^{\otimes k}] \nabla p_z(U)\right).
\end{align}
By triangle inequality, we therefore have
	\begin{align}\label{eq:normgradient}
		\norm{\nabla f_\alpha(U)}_2 \leq \sum_{l=0}^k \norm{ \sum_z A_{\alpha,l,z}(U)}_2 + (k-1) \norm{\sum_z \Tr[X_\alpha R_z(U)^{\otimes k}] \nabla p_z(U)}_2.
	\end{align}
	To evaluate $A_{\alpha,l,z}(U)$ and estimate its norm, let us introduce the operators
	\begin{align}
		B_{\alpha,l,z}(U) =\frac{1}{{d_A}} \Tr_{\overline{l+1}}[X_\alpha R_z(U)^{\otimes l}\otimes\id_{l+1}\otimes R_z(U)^{k-(l+1)}].
	\end{align}
	We then find (using ${U'_\lambda}^\dagger = - U'_\lambda$, since elements of $\mathrm{su(M)}$ are skew-hermitian):
	\begin{align}
		A_{\alpha,l,z}(U_\lambda)^\dagger  = WB_{\alpha,l,z}(U_\lambda)W^\dagger U_\lambda^\dagger \proj{z} 	
- \proj{z} U_\lambda W B_{\alpha,l,z}(U_\lambda)W^\dagger 
	\end{align}
	In particular, this implies 	
	\begin{align}
		\norm{\sum_z A_{\alpha,l,z}(U)}_2 
		&\leq 2 \norm{\sum_z \proj{z} U W B_{\alpha,l,z}(U)W^\dagger}_2 
		= 2\sqrt{\sum_z \bra{z} U W B_{\alpha,l,z}(U)W^\dagger W B_{\alpha,l,z}(U)^\dagger W^\dagger U^\dagger \ket{z}} \notag\\
		&= 2\sqrt{\sum_z \bra{z} U W B_{\alpha,l,z}(U) B_{\alpha,l,z}(U)^\dagger W^\dagger U^\dagger \ket{z}} \notag\\
		&\leq 2\sqrt{\sum_z \norm{W^\dagger U^\dagger \ket{z}}^2 \norm{B_{\alpha,l,z}(U)}^2} \notag\\
		&\leq 2\sqrt{\sum_z p_z(U)d_A{\frac{1}{d_A^{\,2}}}} = {\frac{2}{\sqrt{d_A}}} \label{eq:bound2.1},
	\end{align}
	where we used that $W$ is an isometry, i.e.\ $W^\dagger W=\id_A$, and that $\norm{B_{\alpha,l,z}(U)} \leq {1/d_A}$. Indeed for any normalized vectors $\ket a,\ket b\in \mc H_A$, we have
	\begin{align}
	\begin{split}
		|\bra{a}B_{\alpha,l,z}(U)\ket{b}| 
		&= {\frac{1}{d_A}}|\Tr[X_\alpha R_z(U)^{\otimes l}\otimes \ketbra{b}{a}\otimes R_z(U)^{k-(l+1)}]| \\
		&\leq \frac{\norm{X_\alpha}_2}{{d_A}}\norm{ R_z(U)^{\otimes l}\otimes \ketbra{b}{a}\otimes R_z(U)^{k-(l+1)}}_2\leq \frac{1}{{d_A}},
	\end{split}
	\end{align}
	where we used that $\norm{\ketbra{b}{a}}_2=1=\norm{R_z(U)}_2$.

Turning to the second term in \eqref{eq:normgradient}, we first find
	\begin{align}
		p_z'(U_\lambda) = {\frac{1}{d_A}}\left(\bra{z}U'_\lambda WW^\dagger U_\lambda^\dagger\ket{z} - \bra{z}U_\lambda WW^\dagger U'_\lambda \ket{z}\right).
	\end{align}
	Hence, 
	\begin{align}
		 {d_A} \nabla p_z(U)^\dagger =  WW^\dagger U^\dagger \proj{z} - \proj{z} U WW^\dagger.
	\end{align}
	We thus find that
	\begin{align}
		\norm{\sum_z \Tr[X_\alpha R_z(U)^{\otimes k}]\nabla p_z(U)}_2 &\leq \frac{2}{{d_A}} \norm{\sum_z \Tr[X_\alpha R_z(U)^{\otimes k}] WW^\dagger U^\dagger \proj{z}}_2\notag\\
		&= \frac{2}{{d_A}}\sqrt{\sum_z \left|\Tr[X_\alpha R_z(U)^{\otimes k}]\right|^2 \bra{z}U WW^\dagger U^\dagger\ket{z}} \notag\\
		&= \frac{2}{{d_A}}\sqrt{\sum_z \left|\Tr[X_\alpha R_z(U)^{\otimes k}]\right|^2 d_A p_z(U)}.
	\end{align}
	Since $X_\alpha$ are hermitian and $\norm{X_\alpha}_2 = 1=\norm{R_z(U)^{\otimes k}}_2$, we have $|\Tr[X_\alpha R_z(U)^{\otimes k}]]\leq 1$. Therefore
	\begin{align}\label{eq:bound2.2}
		\norm{\sum_z \Tr[X_\alpha R_z(U)^{\otimes k}]\nabla p_z(U)}_2 &\leq \frac{2}{{\sqrt{d_A}}}.
	\end{align}
	
	Inserting both bounds, 
	\eqref{eq:bound2.1} and \eqref{eq:bound2.2}, 
	into \eqref{eq:normgradient},  we find that the Lipschitz constants are dominated by
	\begin{align}
		L = \max_\alpha \sup_U \norm{\nabla f_\alpha(U)}_2 \leq \frac{2k}{{\sqrt{d_A}}} + \frac{2(k-1)}{{\sqrt{d_A}}} = \frac{2(2k -1)}{{\sqrt{d_A}}}
		\,.
	\end{align}

\end{document}